\documentclass{article}

\usepackage{microtype}
\usepackage{graphicx}
\usepackage{subfigure}
\usepackage{booktabs} 
\usepackage{multirow}
\usepackage{hyperref}

\usepackage[accepted]{icml2025}

\usepackage{amsmath}
\usepackage{amssymb}
\usepackage{enumitem}
\usepackage{mathtools}
\usepackage{amsthm}
\usepackage{colortbl}
\usepackage[capitalize,noabbrev]{cleveref}
\usepackage{pifont}
\theoremstyle{plain}

\theoremstyle{definition}

\theoremstyle{remark}

\definecolor{customgreen}{rgb}{0, 0.69, 0.31}

\usepackage[textsize=tiny]{todonotes}

\icmltitlerunning{Unleashing the Power of Natural Audio Featuring Multiple Sound Sources}

\begin{document}

\twocolumn[
\icmltitle{Unleashing the Power of Natural Audio Featuring Multiple Sound Sources}


\icmlsetsymbol{equal}{*}
\icmlsetsymbol{corr}{$\dagger$}

\begin{icmlauthorlist}
\icmlauthor{Xize Cheng}{zju,equal}
\icmlauthor{Slytherin Wang}{self,equal}
\icmlauthor{Zehan Wang}{zju}
\icmlauthor{Rongjie Huang}{zju}
\icmlauthor{Tao Jin}{zju}
\icmlauthor{Zhou Zhao}{zju,corr}
\end{icmlauthorlist}

\icmlaffiliation{zju}{Zhejiang University, Hangzhou, China}
\icmlaffiliation{self}{Independent Researcher}

\icmlcorrespondingauthor{Xize Cheng}{chengxize@zju.edu.cn}

\icmlkeywords{Sound Separation, Audio-Language Model}

\vskip 0.3in
]



\printAffiliationsAndNotice{\icmlEqualContribution} 

\begin{abstract}

Universal sound separation aims to extract clean audio tracks corresponding to distinct events from mixed audio, which is critical for artificial auditory perception. However, current methods heavily rely on artificially mixed audio for training, which limits their ability to generalize to naturally mixed audio collected in real-world environments. To overcome this limitation, we propose \textbf{ClearSep}, an innovative framework that employs a data engine to decompose complex naturally mixed audio into multiple independent tracks, thereby allowing effective sound separation in real-world scenarios. We introduce two remix-based evaluation metrics to quantitatively assess separation quality and use these metrics as thresholds to iteratively apply the data engine alongside model training, progressively optimizing separation performance. In addition, we propose a series of training strategies tailored to these separated independent tracks to make the best use of them. Extensive experiments demonstrate that ClearSep achieves state-of-the-art performance across multiple sound separation tasks, highlighting its potential for advancing sound separation in natural audio scenarios. For more examples and detailed results, please visit our demo page at \url{https://clearsep.github.io/}.

\end{abstract}

\section{Introduction}

Sound separation~\citep{kavalerov2019universal} aims to separate mixed audio into individual clean tracks. Early researchers focused on speech~\citep{wang2018supervised,wang2023tf} and a limited number of musical instruments~\citep{manilow2022source,luo2023music}, successfully separating these audio sources into individual tracks. Subsequently, some researchers~\citep{yu2017permutation} proposed universal sound separation, attempting to separate various natural sounds.
To further improve sound separation performance, some researchers~\citep{liu2023separate} have proposed using large-scale audio-text datasets to enhance the performance of queried sound separation. Some researchers~\citep{ma2024clapsep,cheng2024omnisep} have even attempted to use multi-modal data as queries to achieve more flexible sound separation.

\begin{figure}
    \centering
    \includegraphics[width=0.7\linewidth]{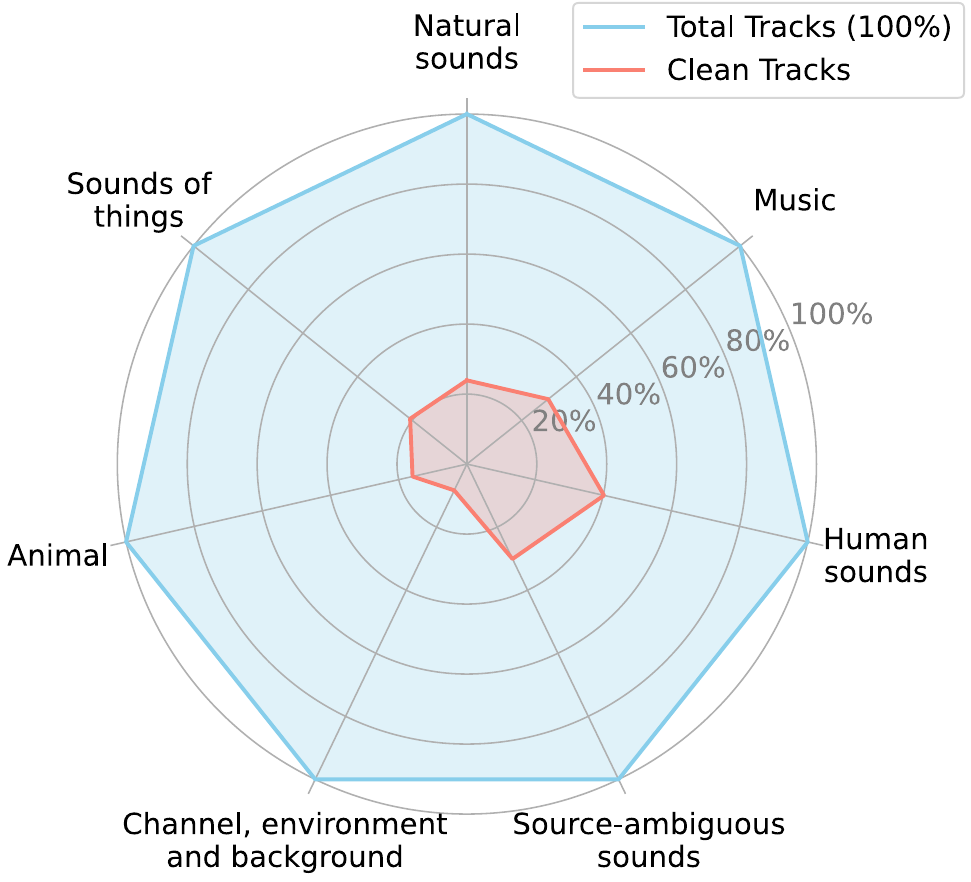}
    \vskip -0.1in
    \caption{Scale comparison between individual clean tracks and total tracks across seven different audio categories in AudioSet. The number of tracks in an audio clip is determined based on the number of audio categories present in AudioSet~\citep{gemmeke2017audio}. \textbf{Clean Tracks} refer to single-source audio containing only a single audio event, while \textbf{Total Tracks} represent the number of independent tracks corresponding to individual audio events extracted from both single-source and mixed-source audio samples.}
    \label{fig:single-class}
    \vskip -0.25in
\end{figure}

Despite these advancements, several challenges remain in existing work: \textbf{(1) Limited applicability to real-world scenarios.} Most existing approaches rely on mixing two highly distinct audio tracks for separation~\citep{wisdom2020unsupervised}, which fails to accurately reflect real-world acoustic conditions. These methods often underperform when faced with mismatched sound types~\citep{manilow2019cutting} or reverberant acoustic environments~\citep{maciejewski2018building}. 
\textbf{(2) Inability to fully utilize natural audio.} Cross-talk contamination during the source recording process often makes it impractical to record both mixture signals and individual ground-truth source signals simultaneously in real-world acoustic environments. Consequently, researchers cannot effectively use naturally mixed audio for training purposes. However, as illustrated in Figure~\ref{fig:single-class}, natural audio contains abundant independent track data that remains unexplored in existing research.
\textbf{(3) Lack of unsupervised methods for evaluating sound separation quality.} In real-world sound separation scenarios, the absence of ground truth makes it difficult to assess the quality of separation results, leading to uncertainty in the reliability of separated tracks.

To address these challenges, we propose a novel sound separation training framework, \textbf{ClearSep}, which leverages a data engine to decompose naturally mixed audio containing multiple sound signals into independent tracks, and adopt these tracks for training more advanced model, unleashing the power of such natural audio.
Specifically, we introduce two unsupervised metrics for assessing sound separation quality: \texttt{Re-SDR} and \texttt{Re-SISDR}. These metrics evaluate the effectiveness of sound separation by remixing the separated independent tracks and measuring the alignment between the remixed audio and the original audio with signal-to-distortion ratio. Then, we use these metrics as thresholds to iteratively extract high-quality independent tracks from naturally mixed audio and train more advanced sound separation models with these tracks in a two-stage loop, continuously optimizing separation performance. Additionally, we propose \texttt{silence augmentation}, which ensures that silence is separated when the target sound signal is absent. This enhances the purity of individual tracks during the data engine process, preventing the introduction of extraneous noise into the separated tracks. 
Finally, to make the best use of these separated independent tracks, we propose two distinct training strategies: \texttt{independent track training} and \texttt{self-separate training}. These strategies pioneer the use of natural audio as training data for mixed audio, significantly improving the practicality for our sound separation model in real-world scenarios.

Experiments on multiple universal sound separation datasets demonstrate that our proposed ClearSep achieves state-of-the-art sound separation performance. To further validate the effectiveness of our model on real-world data, we employed \texttt{Re-SDR} and \texttt{Re-SISDR} to evaluate its performance on natural audio, demonstrating its robustness in real-world scenarios. 
All code and models will be made publicly available. Our main contributions are as follows:
\begin{itemize}[left=0em,itemsep=0pt]
    \vspace{-0.1in}
    \setlength{\leftmargin}{0pt}
    \item We propose leveraging a data engine to separate independent tracks from natural audio and iteratively train more advanced sound separation models in a two-stage process, progressively improving separation performance.
    \item We propose first unsupervised evaluation metric for assessing sound separation quality, enhancing the reliability and consistency of the separated audio tracks.
    \item We propose a series of strategies to make best use of independent tracks extracted from natural audio, enabling training of more advanced sound separation models.
    \item Our ClearSep is the first robust sound separation model tailored for real-world scenarios, achieving SOTA performance across various sound separation benchmarks.

\end{itemize}

\section{Related Works}
\textbf{Universal Sound Separation.} 
Universal sound separation~\citep{kavalerov2019universal} aims to extract distinct audio tracks from mixed audio. Early work mainly addressed domain-specific tasks such as speech separation~\citep{Wang_Chen_2018, pegg2023rtfs} and music separation~\citep{defossez2019demucs, luo2023music}. To extend separation beyond predefined categories, \citet{kavalerov2019universal} introduced permutation invariant training (PIT)~\citep{yu2017permutation}, enabling separation in an unsupervised manner. However, its application was largely restricted to music, speech, and artificial sounds, limiting its effectiveness in real-world scenarios. To achieve truly universal sound separation, MixIT~\citep{wisdom2020unsupervised} leveraged large-scale unlabeled audio data to separate mixed signals into independent tracks corresponding to distinct sound events. Building on this, query-based separation methods~\citep{ochiai2020listen, kong2020source} allowed models to extract specific sound sources based on user-defined queries, further refined by caption-based queries~\citep{liu2023separate} for more fine-grained control.

Despite these advances, current models remain insufficiently robust, particularly in real-world settings, and lack a reliable mechanism to evaluate separation quality without a reference target. To address this, we propose an unsupervised remix-based evaluation method that enhances separation reliability by assessing the consistency between remixed separated tracks and the original audio.

\textbf{Data-Driven Sound Separation.} Research on sound separation has always centered on how to better leverage audio data. Early methods relied on single source audio, employing techniques such as hidden Markov models~\citep{roweis2000one, kristjansson2006super} and non-negative matrix factorization~\citep{smaragdis2004non} to extract distinct signals. Later, synthetic mixtures generated from isolated sources became standard training targets, significantly improving separation performance.
Category-agnostic approaches such as deep clustering~\citep{hershey2016deep} and permutation invariant training (PIT, \citet{yu2017permutation}) enabled speaker-independent speech separation~\citep{hershey2016deep, yu2017permutation} and universal sound separation~\citep{kavalerov2019universal} in various sound categories. Further researchers have focused on scaling datasets~\citep{pons2024gass,liu2023separate} and incorporating fine-grained captions~\citep{ma2024clapsep} to enhance model robustness.

However, most existing methods rely on artificially mixed training data, limiting their ability to generalize to real-world conditions~\citep{wisdom2020unsupervised}. To bridge this gap, we propose ClearSep, an innovative framework that employs a data engine to decompose naturally mixed audio into independent tracks, leading to more robust and effective sound separation in real-world environments.

\begin{figure*}[tb]
    \label{fig:model}
    \begin{center}
    \includegraphics[width=\linewidth]{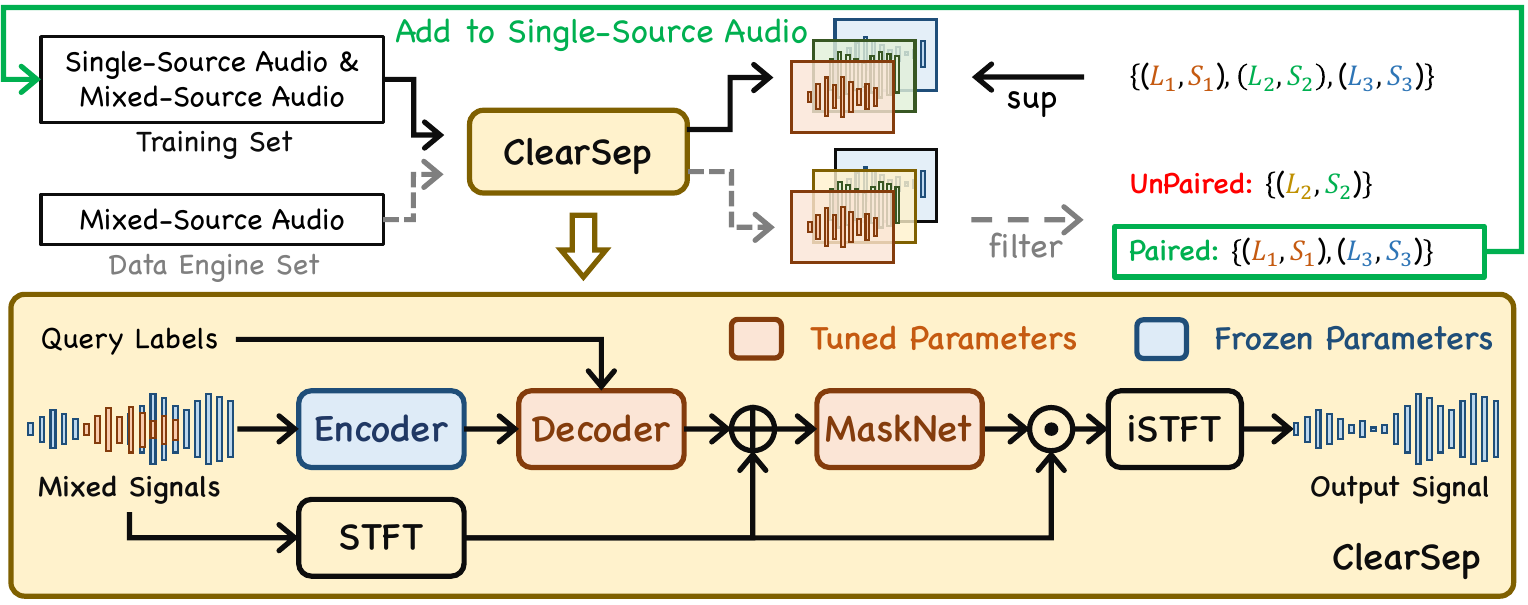}
    \end{center}
    \vskip -0.15in
    \caption{\textbf{Illustration of ClearSep and Data Engine Pipeline.} ClearSep alternates between data engine and model training to progressively enhance sound separation performance and robustness. During the data engine phase, the model employs mutually exclusive class labels as queries to guide separation, ensuring that the separated tracks are independent. A quality filtering process then evaluates the separation results, and only tracks that meet the predefined criteria are incorporated into the single-source audio dataset. In the model training phase, the model is trained with both single-source audio and mixed-source audio, allowing it to achieve more accurate separation.}
    \label{fig:enter-label}
    \vskip -0.15in
\end{figure*}

\section{Methods}

\subsection{Overview}
Universal sound separation focuses on isolating specific audio signals from a mixture based on a given query. Traditional methods typically artificially mixed two distinct audio tracks and adopted one as the target for training, as described in Section~\ref{sec:sound_separate}. In this work, we further propose a data engine for sound separation, which works iteratively with model training to progressively enhance the performance. As described in Section~\ref{sec:data-engine}, each complex audio $X$ containing multiple distinct sound events $\{{T}_1, \dots, {T}_N\}$ can be separated into independent tracks $\{{X}_1, \dots, {X}_N\}$, providing cleaner and more event-specific training data. Each track ${X}_i$ corresponds to a specific sound event $T_i$.

\subsection{Learning from Artificially Mixed Audio Tracks}
\label{sec:sound_separate}


ClearSep adopts an architecture similar to CLAPSep~\citep{ma2024clapsep}, as illustrated in Figure~\ref{fig:model}. It uses the pre-trained CLAP model to extract query embeddings $\mathbf{Q}$ and hierarchical audio features $\{\mathbf{H}^0_{e}, \dots, \mathbf{H}^L_{e}\}$ from the input mixture $\tilde{X}$. The decoder processes these audio features, integrating them with $\mathbf{Q}$ through a U-Net structure to iteratively refine the target sound representation $\mathbf{H}_t$. MaskNet, composed of transformer encoder layers, predicts a spectrogram mask that is applied to the magnitude spectrogram of the mixture. Finally, the clean separated audio $\hat{X}$ is reconstructed using the inverse short-time Fourier transform (ISTFT), combining the estimated mask with the phase of the original mixture $\tilde{X}$, thus ensuring high-fidelity sound separation.

\textbf{Query Embedding.} We concatenate the positive embedding $\mathbf{Q}_{\text{p}} \in \mathbb{R}^{D}$, extracted from the positive query corresponding to the target audio, with the negative embedding $\mathbf{Q}_{\text{n}} \in \mathbb{R}^{D}$, representing the interfering audio, to form the conditional embedding $\mathbf{Q} \in \mathbb{R}^{2D}$. During the training, we employ three query settings with proportions of 0.25, 0.25, and 0.5, respectively: \textit{pos-only}, \textit{neg-only}, and \textit{pos+neg}. If either the positive or negative query is missing, the corresponding embedding is set to $\mathbf{0} \in \mathbb{R}^{D}$.

\textbf{Audio Encoder.} 
The mel-spectrogram ${|\tilde{X}|_{mel}} \in \mathbb{R}^{T \times F}$ of the sound mixture $\tilde{X}$ is reshaped into a patch sequence $\mathbf{H}^0_{e} \in \mathbb{R}^{\left(\frac{T}{P} \times \frac{F}{P}, D_f\right)}$, where $T$ and $F$ denote the time frames and frequency bins, $P$ is the patch size, and $D_f$ is the feature dimension. This sequence is processed through the $L$-layer CLAP encoder, producing $L$ layer-wise audio features, ${\{\mathbf{H}^0_{e}, \dots, \mathbf{H}^L_{e}\}}$. Each layer (e.g. $l$-th layer) applies a function $f_{\text{enc}}^l(\cdot)$, which includes a patch-merging module that halves the time and frequency dimensions of $\mathbf{H}^{l-1}_e$ while doubling the feature dimension:
\begin{equation}
    \mathbf{H}^l_{e} = f_{\text{enc}}^l(\mathbf{H}^{l-1}_{e})
\end{equation}
\textbf{Decoder.}\quad 
The decoder follows the U-Net structure to extract the target sound feature $\mathbf{H}_t$ by hierarchically aggregating layer-wise audio features and conditional embedding $\mathbf{Q}$. The \(l\)-th decoder feature \(\mathbf{H}_d^l\) is obtained with \(l\)-th decoder layer \(f_{\text{dec}}^l(\cdot)\) and two linear layers \(\gamma^{L-l}(\cdot)\) and \(\beta^{L-l}(\cdot)\):
\begin{equation} 
\mathbf{H}_d^l = f_{\text{dec}}^l(\mathbf{H}^{L-l}_d) + \gamma^{L-l}(\mathbf{Q}) \mathbf{H}^{L-l}_{e} + \beta^{L-l}(\mathbf{Q}),
\end{equation}
where the initial decoder input \(\mathbf{H}^0_d\) is set to the final encoder feature \(\mathbf{H}^L_e\). Each decoder layer is followed by a patch-expanding module, which increases the dimensions of tokens and patches while reducing their feature dimensions, ensuring alignment between \(\mathbf{H}^l_d\) and the corresponding encoded features \(\mathbf{H}^{L-l}_e\). Finally, an inverse patch-embedding process, including a transpose convolution operation, is applied, followed by reshaping to produce the final target sound feature \(\mathbf{H}_t \in \mathbb{R}^{T \times F}\).

\textbf{MaskNet.}\quad MaskNet, consisting of $N$ transformer encoder layers, takes as input the concatenation of the target sound feature $\mathbf{H}_t$ and the linear magnitude spectrogram $|\tilde{X}|$ of the sound mixture $\tilde{X}$ to generate a spectrogram mask $\mathbf{M}$. Specifically, an STFT module is used to compute the magnitude and phase spectrograms of the sound mixture:
\begin{equation}
\text{STFT}(\tilde{X}) = \tilde{X} = |\tilde{X}| e^{j\Phi_{\tilde{X}}},
\end{equation}
where $\tilde{X}$ is the complex spectrogram, 
$|\tilde{X}| \in \mathbb{R}^{T \times F}$ is the magnitude spectrogram, $\Phi_{\tilde{X}} \in \mathbb{R}^{T \times F}$ denotes the phase spectrogram, and $j$ is the imaginary unit.
The mask $\mathbf{M}$ is activated by Sigmoid activation to constrain its values to $(0, 1)$. Finally, the target sound waveform $\hat{X}$ is reconstructed using the inverse short-time Fourier transform (ISTFT):
\begin{equation}
\hat{X} = \text{ISTFT}(\mathbf{M} \odot |\tilde{X}| e^{j\Phi_{\tilde{X}}}),
\end{equation}
where $\odot$ denotes element-wise multiplication, and $\Phi_{\tilde{X}}$ is reused from the sound mixture as an estimate for the phase of the extracted sound source.

\textbf{Training Objective.} \quad
During training, the Audio Encoder is frozen and fine-tuned with LoRA~\citep{hu2021lora}. The model is optimized with SDR and SISDR:
\begin{equation}
    L(\hat{x}, x) = -\lambda \, \text{SDR}(\hat{x}, x) - (1 - \lambda) \, \text{SISDR}(\hat{x}, x),
\end{equation}
where $\lambda = 0.9$ is set based on prior work~\citep{ma2024clapsep}.


\subsection{Unleashing the Power of Natural Audio Featuring Multiple Sound Sources}
\label{sec:data-engine}

Natural sounds often consist of multiple distinct audio events. To fully utilize these audio samples, it is essential to separate authentically mixed audio tracks into independent tracks corresponding to each event. We propose a data engine strategy for sound separation that iteratively alternates between leveraging pre-trained sound separation models to extract high-quality independent tracks and training more advanced separation models using these tracks.

\subsubsection{Separating Independent Tracks}

Modern audio event recognization works~\citep{kong2020panns,chen2025slam} has made it possible to effectively identify various types of audio events. For a given natural audio $X$, we can determine its $m$ distinct audio event categories $\{T_1, …, T_m\}$\footnote{For AudioSet, we directly use its provided category labels.}. With a well-performing sound separation model $\mathcal{S}$, we can separate ${X}$ into $m$ independent tracks $\{(T_1,\hat{X}_1),...,(T_m,\hat{X}_m)\}$ corresponding to different events. Please note that audio-event labels can serve as negative queries for one another in the sound separation process. The combination of positive and negative queries (pos+neg), which encompasses comprehensive information, enables the separation of cleaner sound signals and effectively guides the training of models that rely probabilistically on positive queries (\textit{pos-only}) or negative queries (\textit{neg-only}) in training sound separation models.

\textbf{Silence Augmentation.} Existing studies often prioritize separating the target signal while overlooking the importance of avoiding unrelated signals. This oversight can result in separated tracks being contaminated with inconsistent or extraneous audio. To address this, we propose a method called silence augmentation, designed to ensure that the separated tracks are silent when the given audio does not contain sounds associated with the target label. During training, we randomly select sound events that are absent from $\tilde{X}$ as queries and train the model to output no content in such cases (i.e., silence, $\mathbf{0}$).

\textbf{Remix-based Filter.} To improve training quality, we propose a filtering mechanism that selectively incorporates high-quality separated tracks into the training dataset. When the separation performs well, the separated tracks should contain no overlapping signals, and the re-mixed audio generated from these independent tracks should closely match the original audio. Based on this observation, we introduce two remix-based metrics as thresholds to filter qualified data for subsequent training. Specifically, Re-SDR and Re-SISDR are used to measure the signal-to-distortion ratio~\citep{vincent2006performance} and the scale-invariant signal-to-distortion ratio~\citep{le2019sdr}, respectively, between the natural audio $X$ and the re-mixed audio $\bar{X}$. For more details on Re-SDR and Re-SISDR, please refer to Appendix~\ref{app:remix}.



\subsubsection{Training with Separated Tracks.}

With these separated independent tracks, we propose two methods for training advanced sound separation models.

\textbf{Independent Track Training.}
Any separated track pair $(T_i, \hat{X}_i)$ can be treated as a clean track and mixed with other tracks, using $\hat{X}_i$ as the target for training. Since the mapping relationship between $T_i$ and $\hat{X}_i$ is more explicit within independent tracks, this approach effectively clarifies the association between each audio event and its signal.

\textbf{Self-Separate Training.}
To better adapt to real-world data, we propose Self-Separate Training (SST), which treats natural audio $X$ as mixture audio $\tilde{X}$ and an independent track $\hat{X}_i$ as the target for training. The positive query is defined as $T_i$, while the negative queries are set to $\{{T_1, \cdots, T_{i-1}, T_{i+1}, \cdots, T_m}\}$.


\begin{table*}[]
\caption{Comparison of sound separation performance across different models. \textbf{\textsc{Audio}} represents duration of audio data used for training. \textbf{\textsc{Lab}} denotes label-based data, while \textbf{\textsc{Cap}} refers to caption-based data. Experiments marked with $^\dagger$ are taken from~\citet{ma2024clapsep}.} 
\label{tab:main}
\setlength\tabcolsep{3pt}
\begin{center}
\begin{sc}
\begin{small}
\begin{tabular}{@{}lrrcccccc@{}}
\toprule
\multicolumn{1}{l}{\multirow{2}{*}{\textbf{Methods}}} & \multicolumn{2}{c}{\textbf{Audio(h)}} & \multicolumn{2}{c}{\textbf{AudioCaps}} & \multicolumn{2}{c}{\textbf{AudioSet}} & \multicolumn{2}{c}{\textbf{ESC-50}} \\ \cmidrule(l){4-9} 
\multicolumn{1}{c}{}                       
& \textbf{Lab}
& \textbf{Cap}
& \textbf{SDRi}    
& \textbf{SISDRi}   
& \textbf{SDRi}   
& \textbf{SISDRi}    
& \textbf{SDRi}  
& \textbf{SISDRi}   \\ \midrule
\multicolumn{9}{l}{\cellcolor[HTML]{EFEFEF}\textit{Querying with Only Positive Query.}}                                                                                                  \\
LASS\scriptsize{~\citep{liu2022separate}}              
& -  
& 17 
& 0.33$\pm$7.39        
& -1.11$\pm$8.18       
& 2.57$\pm$7.86      
& 4.41$\pm$8.60    
& -0.48$\pm$9.97    
& -2.60$\pm$11.31       \\
LASS$^\dagger$\scriptsize{~\citep{liu2022separate}}                                                
& -   
& 145 
& 6.75$\pm$5.59    
& 6.05$\pm$5.86      
& 3.12$\pm$7.61      
& 2.02$\pm$8.29            
& 7.49$\pm$9.06      
& 6.07$\pm$10.18        \\
AudioSep\scriptsize{~\citep{liu2023separate}}   
& 5\,800
& 8\,300
& 7.75$\pm$5.59        
& 7.04$\pm$5.72       
& 8.02$\pm$6.23       
& 7.26$\pm$6.44            
& 10.33$\pm$7.61     
& 9.20$\pm$7.97         \\
AudioSep$^\dagger$\scriptsize{~\citep{liu2023separate}} 
& - 
& 145 
& 6.63$\pm$5.46      
& 5.55$\pm$5.77        
& 3.81$\pm$6.64       
& 2.30$\pm$7.29        
& 7.66$\pm$7.92     
& 5.81$\pm$8.78         \\
ClapSep\scriptsize{~\citep{ma2024clapsep}}  
& -
& 145      
& 9.64$\pm$5.09         
& 8.92$\pm$5.27         
& 8.02$\pm$7.17       
& 7.05$\pm$7.60                  
& 12.23$\pm$7.52    
& 11.14$\pm$8.01        \\
ClapSep\scriptsize{~\citep{ma2024clapsep}}           
& 5\,800      
& -
& 9.83$\pm$5.70                
& 9.19$\pm$5.95                   
& 9.56$\pm$6.77                
& 8.91$\pm$7.18                  
& 12.70$\pm$7.62                
& 11.84$\pm$8.17     \\
\rowcolor[HTML]{DDEBF7}
ClearSep(ours)                                         
& 5\,800          
& -
& 10.33$\pm$5.49               
& 9.70$\pm$5.69                 
& \textbf{10.45$\pm$6.40}            
& \textbf{9.74$\pm$6.71}           
& 13.64$\pm$7.24              
& 12.61$\pm$7.61                 \\
\rowcolor[HTML]{DDEBF7}
ClearSep(ours)                                  
& 5\,800    
& 145
& \textbf{10.55$\pm$5.22}          
& \textbf{9.93$\pm$5.44}              
& 10.31$\pm$6.53               
& 9.57$\pm$6.90                 
& \textbf{13.64$\pm$6.91}             
& \textbf{12.78$\pm$7.16}                  \\
\midrule
\multicolumn{9}{l}{\cellcolor[HTML]{EFEFEF}\textit{Querying with Positive Query and Negative Query.}}                                                                                                  \\
ClapSep\scriptsize{~\citep{ma2024clapsep}}                     
& -   
& 145 
& 10.08$\pm$4.42     
& 9.40$\pm$4.45     
& 9.29$\pm$5.61      
& 8.44$\pm$5.75      
& 13.09$\pm$6.22   
& 12.10$\pm$6.37        \\
ClapSep\scriptsize{~\citep{ma2024clapsep}}  
& 5\,800 
& -
& 10.42$\pm$4.47                 
& 9.83$\pm$4.51                  
& 10.63$\pm$5.17                
& 9.97$\pm$5.27                 
& 13.44$\pm$6.43             
& 12.56$\pm$6.54                  \\
\rowcolor[HTML]{DDEBF7}
ClearSep(ours)  
& 5\,800  
& -         
& 10.93$\pm$4.48                
& 10.33$\pm$4.48              
& \textbf{11.23$\pm$5.07}            
& \textbf{10.58$\pm$5.14}           
& \textbf{14.16$\pm$6.29}              
& 13.20$\pm$6.34                  \\
\rowcolor[HTML]{DDEBF7}
ClearSep(ours)  
& 5\,800 
& 145      
& \textbf{10.97$\pm$4.42}
& \textbf{10.39$\pm$4.43}                  
& 11.14$\pm$5.05                
& 10.48$\pm$5.14                 
& 14.06$\pm$6.19              
& \textbf{13.26$\pm$6.23}                  \\
\bottomrule
\end{tabular}
\end{small}
\end{sc}
\end{center}
\vskip -0.2in
\end{table*}

\section{Experiments}

\subsection{Implementation Details}


Our ClearSep builds upon CLAPSep~\citep{ma2024clapsep}, utilizing CLAP~\citep{laionclap2023} as the audio encoder and exclusively employing text-based queries during training. We process 10-second audio samples at a 32kHz sampling rate. The model is trained with a learning rate of 2e-4 and a batch size of 256 on four 3090 GPUs. For more details on training implementation, please refer to Appendix \ref{app:training}.

In the data engine process, we enforce strict quality criteria for training samples used in Self-Separate Training (SST) and Iterative Track Training (ITT). Specifically, for a sample to be included in SST, it must satisfy both Re-SDR $> 15\,$dB and Re-SISDR $> 15\,$dB, ensuring that only high-purity separations contribute to self-supervised learning. Meanwhile, samples used for ITT must meet a threshold of Re-SDR $> 10\,$dB and Re-SISDR $> 10\,$dB, allowing for a broader range of training data while maintaining sufficient separation quality. During the first iteration, we train the model for 80 epochs. In subsequent iterations, the model undergoes fine-tuning for 20 additional epochs, using the checkpoint from the previous iteration. 
For further details on the data engine process, please refer to Appendix~\ref{app:data_engine}.

For comparison, we conducted training on both AudioSet and AudioCaps at varying dataset scales, with evaluation performed across three general sound datasets: AudioCaps, AudioSet, and ESC-50~\citep{piczak2015esc}. Following ~\citet{ma2024clapsep}, our primary evaluation metrics are SDRi and SISDRi, supplemented by SDR and SISDR measurements for additional experimental reference. Please note that due to some overlap between the AudioCaps test set and the AudioSet training set, we removed the overlapping data from the AudioSet training set to ensure a fair comparison.

\subsection{Main Results}

Since our proposed unsupervised sound separation quality evaluation requires that separated tracks be mutually exclusive and collectively exhaustive, the sound separation data engine is built on label-queried separation. In Table~\ref{tab:main}, we compare the advantages and limitations of label-queried and caption-queried sound separation and evaluate the performance improvements achieved through the data engine.

\subsubsection{Label-Queried vs. Caption-Queried}

Compared to caption-queried audio data, label-queried audio data is easier to annotate and highly scalable, making it more effective for expanding training datasets. When training CLAPSep with 5,800 hours of label-queried data, it significantly outperforms the model trained on only 145 hours of caption-queried data, achieving an SDRi improvement from 8.02 dB to 9.56 dB (+1.54 dB) on AudioSet. Notably, label-trained CLAPSep not only excels on label-queried datasets but also demonstrates superior performance on AudioCaps, where open-vocabulary captions serve as queries. Specifically, the label-trained CLAPSep achieves a 0.19 dB SDRi improvement (from 9.64 dB to 9.83 dB) over its caption-trained counterpart, reinforcing the notion that dataset scale plays a more critical role than the granularity of query annotations.

Furthermore, in Appendix \ref{app:lab_cap}, we evaluate the performance of models trained on label-queried and caption-queried data under the same dataset scale for label-based sound separation tasks. The results indicate that when querying with labels, models trained on label-queried data outperform those trained on caption-queried data, achieving an SDRi improvement from 8.16 dB to 8.41 dB (+0.25 dB). This demonstrates that while caption-queried data help to achieve open-vocabulary sound separation, it is less effective when the task requires separation based on a fixed set of predefined labels. In contrast, label-queried training allows the model to better capture target audio signals, leading to more robust and precise sound separation in scenarios where labels serve as structured queries.

Moreover, label-based queries provide practical advantages by eliminating the need for manually crafted captions, enabling a more comprehensive construction of positive and negative queries. Our experiments show that models trained with both positive and negative queries significantly outperform those trained with positive-only queries. For example, on AudioSet, ClearSep achieves a SDRi improvement from 10.45 dB to 11.23 dB (+0.78 dB), while its performance stability is enhanced, as indicated by a reduction in standard deviation from 6.40 dB to 5.05 dB. These results further highlight the advantages of label-queried sound separation, demonstrating its effectiveness in producing cleaner and more stable separation outcomes.

\begin{table}[]
\caption{Comparison of ClearSep Across Different Iterations on AudioSet. \textbf{\textsc{Tracks}} represents the total duration after applying the data engine for sound separation.}
\label{tab:iter}
\vskip -0.2in
\begin{center}
\begin{small}
\begin{sc}
\begin{tabular}{@{}lrcc@{}}
\toprule
\textbf{Methods} & \textbf{Tracks} & \textbf{SDRi} & \textbf{SISDRi}         \\ \midrule
ClearSep Iter1   & 5\,800h     & 9.56$\pm$6.77            & 8.91$\pm$7.18           \\
ClearSep Iter2   & 9\,429h     & 10.25$\pm$6.50            & 9.54$\pm$6.83           \\
ClearSep Iter3   & 10\,288h     & \textbf{10.45$\pm$6.40}  & \textbf{9.74$\pm$6.71} \\ \bottomrule
\end{tabular}
\end{sc}
\end{small}
\end{center}
\vskip -0.2in
\end{table}

\subsubsection{Data Engine for Sound Separation} 



Leveraging the data engine to fully exploit mixed-source audio enables the training of more advanced sound separation models. When trained on the same dataset scale (5,800 hours from AudioSet), ClearSep demonstrates superior sound separation performance. Specifically, on ESC-50, SDRi improves from 12.70 dB to 13.64 dB (+0.94 dB), while also exhibiting greater stability in performance.

In Appendix \ref{app:clean}, we compare training on single-source audio versus mixed-source audio. The results indicate that models trained on single-source data consistently outperform those trained on mixed-source data, with SDRi increasing from 8.72 dB to 9.16 dB (+0.44 dB). This highlights the importance of decomposing mixed-source audio into independent tracks to enhance sound separation models.

Additionally, in Table \ref{tab:iter}, we present a comparative analysis of ClearSep’s performance across different iterations. As the iterative process progresses, the amount of separated multi-source audio continues to expand, driving gradual yet consistent improvements in model performance. SDRi increases from 9.56 dB to 10.45 dB (+0.89 dB), providing strong evidence of the effectiveness of ClearSep in achieving high-quality sound separation.

\subsection{Purity Experiment of Sound Separation}

\begin{table}[]
\caption{Comparison of sound separation purity on AudioCaps. Evaluating whether the model can separates silence when the given audio does not contain the target source. ({P}): Querying with a Pos-Only query. ({P+N}): Querying with Pos and Neg queries.}
\label{tab:silence}
\begin{center}
\begin{small}
\begin{sc}
\begin{tabular}{@{}lcc@{}}
\toprule
\textbf{Methods}                                  & \textbf{Silence-SDR}      & \textbf{Silence-SISDR}\\ \midrule
\scriptsize{AudioSep}                                     
& 6.65$\pm$3.85                
& 5.65$\pm$4.94   \\
\scriptsize{ClapSep(P+N)}                             
& 6.63$\pm$2.85                 
& 2.21$\pm$2.00   \\
\rowcolor[HTML]{DDEBF7}
\scriptsize{ClearSep(P)} 
& 57.37$\pm$61.99   
& 41.53$\pm$47.81  \\
\rowcolor[HTML]{DDEBF7}
\scriptsize{ClearSep(P+N)} 
& \textbf{107.09$\pm$53.96} 
& \textbf{78.96$\pm$41.10}  \\ \bottomrule
\end{tabular}
\end{sc}
\end{small}
\end{center}
\vskip -0.2in
\end{table}

In sound separation, minimizing unwanted noise unrelated to the target source is essential to achieve high-quality separation. Following previous research~\citep{wang2024consistent}, we adopt Silent-SDR and Silent-SISDR to assess the purity of separated tracks. These metrics are specifically designed to determine whether the model correctly avoids extracting any signal (i.e., a silence signal, $\mathbf{0}$) when the given audio does not contain sounds corresponding to the target source. For further details on Silent-SDR and Silent-SISDR, please refer to Appendix~\ref{app:silence}.

As shown in Table~\ref{tab:silence}, we assess the separation purity by evaluating whether the model introduces interfering noise when queried with labels that are absent from the mixed audio in the AudioCaps dataset. Without silence augmentation, AudioSep and CLAPSep demonstrate suboptimal Silent-SDR performance, achieving only 6.63 dB, which indicates significant interference from irrelevant noise. In contrast, ClearSep, with silence augmentation, achieves a substantial improvement, boosting Silent-SDR to 57.37 dB, effectively suppressing unrelated noise. Furthermore, when querying with both positive and negative queries (P+N), the separation purity is further enhanced, reaching 107.09 dB. These results highlight ClearSep superior capability in achieving cleaner and more precise separations.

\begin{table}[]
\caption{Comparison of Sound Separation Performance on Real-World Data. Evaluating sound separation performance on mixed-source natural audio from the AudioSet test set using Re-SDR and Re-SISDR. $\textsc{Prop}_{>15\,\mathrm{dB}}$ represents the proportion of samples where the corresponding metric exceeds 15 dB.}
\label{tab:real}
\setlength\tabcolsep{3pt}
\begin{center}
\begin{small}
\begin{sc}
\begin{tabular}{@{}lcccc@{}}
\toprule
\multirow{2}{*}{\textbf{Methods}} & \multicolumn{2}{c}{\textbf{Re-SDR}} & \multicolumn{2}{c}{\textbf{Re-SISDR}} \\ \cmidrule(l){2-5} 
                                  & Mean               & $\text{Prop}_{>15\,\mathrm{dB}}$            & Mean                & $\text{Prop}_{>15\,\mathrm{dB}}$             \\ \midrule
\scriptsize{AudioSep}                       
& 14.679                  
& 44\%              
& 12.297                   
& 27\%                \\
\scriptsize{CLAPSep(P+N)}                     
& 16.764          
& 62\%           
& 14.629             
& 42\%             \\
\rowcolor[HTML]{DDEBF7}
\scriptsize{ClearSep(P+N)}                          & \textbf{18.424}             & \textbf{73\%}            & \textbf{16.164}              & \textbf{57\%}             \\ \bottomrule
\end{tabular}
\end{sc}
\end{small}
\end{center}
\vskip -0.2in
\end{table}

\subsection{Sound Separation on Real World Data}

As shown in Table~\ref{tab:real}, we evaluate the sound separation performance on real-world mixed-source audio by decomposing naturally mixed recordings into independent tracks, remixing them, and assessing the separation quality using Re-SDR and Re-SISDR. The results demonstrate that our proposed ClearSep significantly outperforms both AudioSep and CLAPSep, achieving the best performance across Re-SDR and Re-SISDR. Specifically, ClearSep achieves an Re-SDR of 18.424 dB, surpassing ClearSep (16.764 dB) and AudioSep (14.679 dB), showcasing its superior ability in real-world sound separation. Additionally, 73\% of the separated samples exceed 15 dB in Re-SDR, indicating that the majority of tracks maintain high fidelity. These improvements stem from our data engine approach, which effectively utilizes naturally mixed-source audio for training, and our self-separate training, which enhances model robustness on real-world samples.

\begin{table}[h]
\caption{Comparison of Silence Augmentation with different silence rates (\textbf{\textsc{SIL RATE}}) on AudioCaps. Silence rates $\alpha$ represent the proportion of silence samples included during training.}
\label{tab:silence}
\setlength\tabcolsep{3pt}
\begin{center}
\begin{small}
\begin{sc}
    \begin{tabular}{@{}ccccc@{}}
    \toprule
    \textbf{SIL Rate}                        & \textbf{SDR}           & \textbf{SDRi}          & \textbf{SI-SDR}        & \textbf{SI-SDRi}       \\ \midrule
    $\alpha=0.00$  & 9.09$\pm$5.16          & 9.07$\pm$5.16          & 8.36$\pm$5.29          & 8.36$\pm$5.29          \\
    $\alpha=0.05$  & \textbf{9.23$\pm$5.23} & \textbf{9.21$\pm$5.23} & \textbf{8.49$\pm$5.36} & \textbf{8.49$\pm$5.36} \\
    $\alpha=0.10$  & 9.13$\pm$5.30          & 9.11$\pm$5.30          & 8.39$\pm$5.42          & 8.39$\pm$5.42          \\
    $\alpha=0.15$  & 9.11$\pm$5.24          & 9.09$\pm$5.24          & 8.38$\pm$5.36          & 8.37$\pm$5.36          \\ \bottomrule
    \end{tabular}
\end{sc}
\end{small}
\end{center}
\vskip -0.2in
\end{table}

\subsection{Ablation Studies}
\subsubsection{Silence Augmentation}
To evaluate the effectiveness of silence augmentation and determine its optimal configuration, we systematically trained models with varying silence rates $\alpha$, as summarized in Table~\ref{tab:silence}. Our findings indicate that incorporating silence augmentation consistently improves separation performance, with SDRi ranging from 9.11 dB to 9.21 dB, surpassing the baseline model without silence augmentation ($\alpha=0$) by 0.02–0.12 dB. This enhancement suggests that silence augmentation strengthens the model’s ability to distinguish target signals from interfering noise, leading to cleaner and more precise audio separation. The observed benefits stem from the model’s improved capacity to suppress false positives in silent regions, effectively reducing noise leakage between separated tracks.

However, as the silence rate increases to $\alpha=0.10$ or $\alpha=0.15$, performance begins to degrade, with SDRi dropping by 0.10 dB and 0.12 dB, respectively, compared to the optimal $\alpha=0.05$ setting. This decline highlights the necessity of maintaining a balanced training set, ensuring that sufficient paired audio samples remain to support robust sound separation learning. While silence augmentation proves beneficial, our findings suggest that $\alpha=0.05$ is the most suitable silence rate, offering an optimal trade-off between improving separation quality and maintaining a diverse training distribution without disproportionately emphasizing silent regions at the expense of meaningful sound separation learning.

\begin{table}[]
\caption{Performance comparison of training with filtered samples at different thresholds on AudioCaps. \textbf{\textsc{ITT}} stands for Independent Track Training, while \textbf{\textsc{SST}} represents Self-Separation Training. \textbf{\# \textsc{Audio(h)}} indicates the scale of training audio available after processing with the Data Engine, with (the numbers in parentheses) denoting the subsets specifically suitable for SST.}
\label{tab:data-centric}
\setlength\tabcolsep{2pt}
\begin{center}
\begin{small}
\begin{sc}
\begin{tabular}{@{}lcclcc@{}}
\toprule
\multirow{2}{*}{\textbf{ID}} & \multicolumn{2}{c}{\textbf{Threshold}} & \multirow{2}{*}{\textbf{\#Audio(h)}} & \multirow{2}{*}{\textbf{SDRi}} & \multirow{2}{*}{\textbf{SI-SDRi}} \\ \cmidrule(lr){2-3}
                             & \textbf{ITT}       & \textbf{SST}      &                              &                                &                                   \\ \midrule
$\textit{R}_1$                   
& -            
& -  
& 145(0)             
& 8.98$\pm$4.93       
& 8.24$\pm$5.01        \\ \midrule
$\textit{R}_2$        
& $>10\,\mathrm{dB}$               
& -            
& 298(0)               
& \textbf{9.29$\pm$5.18}        
& \textbf{8.54$\pm$5.28}        \\
$\textit{R}_{3}$          
& $>10\,\mathrm{dB}$                  
& $>10\,\mathrm{dB}$   
& 145(67)                    
& 9.24$\pm$5.23        
& 8.45$\pm$5.38        \\ \midrule
$\textit{R}_4$          
& $>15\,\mathrm{dB}$                     
& -            
& 212(0)               
& 9.18$\pm$5.16       
& 8.43$\pm$5.26        \\
$\textit{R}_{5}$          
& $>15\,\mathrm{dB}$                       
& $>15\,\mathrm{dB}$       
& 145(29)               
& \textbf{9.30$\pm$5.15}        
& \textbf{8.53$\pm$5.26}        \\ \midrule
$\textit{R}_{6}$          
& $>10\,\mathrm{dB}$                     
& $>15\,\mathrm{dB}$        
& 298(29)               
& \textbf{9.32$\pm$5.19}        
& \textbf{8.56$\pm$5.29} \\ \bottomrule
\end{tabular}
\end{sc}
\end{small}
\end{center}
\vskip -0.3in
\end{table}

\subsubsection{Data-Centric Sound Separation}

After separating multi-label mixed audio into individual tracks using a sound separation model, we incorporated the filtered high-quality independent tracks into the training process to develop a data-centric approach to sound separation. To evaluate the effectiveness of this data engine, we conducted experiments under different threshold settings and analyzed their impact on model performance, as shown in Table~\ref{tab:data-centric}. Specifically, we experimented with two threshold levels, 10 dB and 15 dB, and trained models using two distinct methods: Independent Track Training (ITT) and Self-Separation Training (SST).

ITT improves model performance by increasing the diversity of training samples, where larger training datasets play a more significant role in enhancing effectiveness. For instance, in $\textit{R}_{2}$, adopting a 10 dB threshold resulted in 86 additional hours of training data compared to the 15 dB threshold, leading to a 0.11 dB SDRi improvement, highlighting the advantage of expanding data diversity.
In contrast, SST benefits more from high-quality separated tracks, as it relies on self-separation and requires ensuring that the query does not correspond to other signals in the original audio. As a result, in $\textit{R}_{5}$, using a 15 dB threshold led to a 0.06 dB SDRi improvement over $\textit{R}_{3}$, confirming that higher-quality training data is particularly beneficial for SST.

To further optimize performance, we applied a setting $\textit{R}_{6}$, leveraging both ITT and SST with their respective optimal data subsets. This approach achieved the best performance, reaching an SDRi of 9.51, marking a 0.53 dB improvement over $\textit{R}_{1}$. These results validate the effectiveness of our data-centric sound separation framework, demonstrating that the iterative refinement of naturally mixed audio significantly enhances model robustness and separation quality.

\begin{table*}[]
\caption{Qualitative Results of Sound Separation with Non-Existing Queries. A non-existing query is used as the positive query, while the actual labels of the audio serve as negative queries. To ensure consistency in visualization, all Mel spectrograms maintain a consistent dynamic range on the logarithmic scale.}
\label{tab:silence_image}
\vskip 0.1in
\begin{center}
\begin{small}
\begin{sc}
\begin{tabular}{@{}ccccc@{}}
\toprule
\textbf{Query} & \textbf{Mixture} & \textbf{AudioSep} & \textbf{CLAPSep(P+N)} & \textbf{ClearSep(P+N)} \\ \midrule
\includegraphics[width=0.18\textwidth]{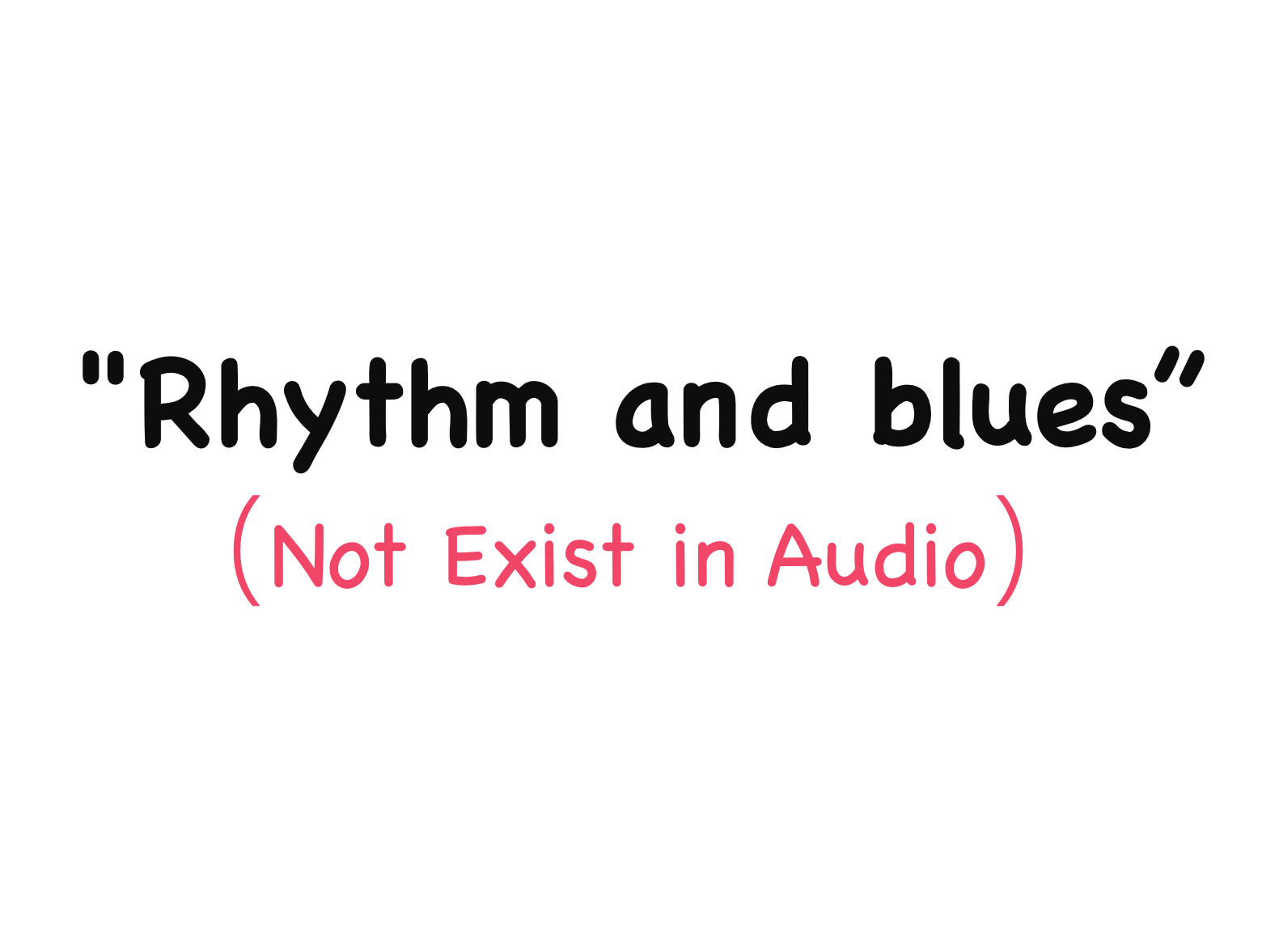}         
&\includegraphics[width=0.18\textwidth]{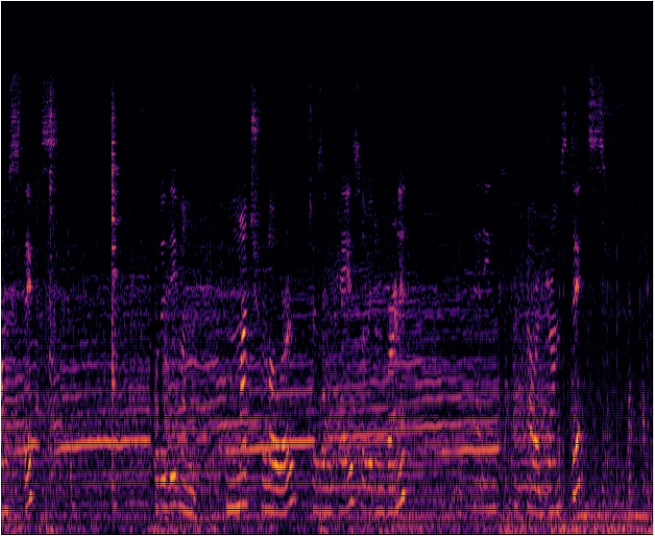}         
&\includegraphics[width=0.18\textwidth]{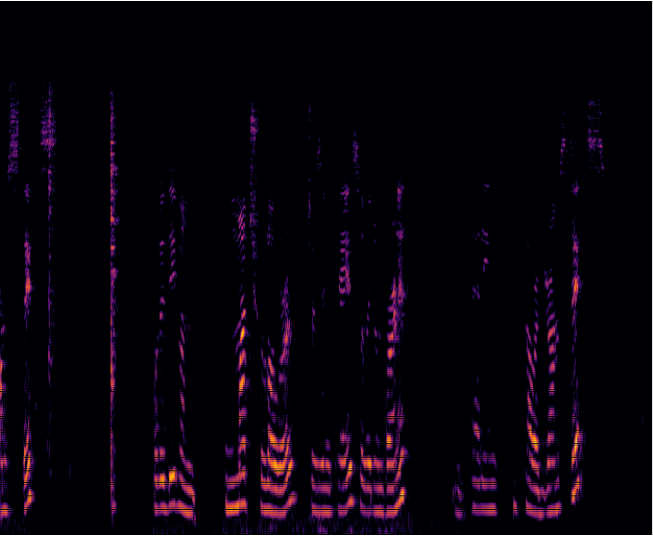}                   
&\includegraphics[width=0.18\textwidth]{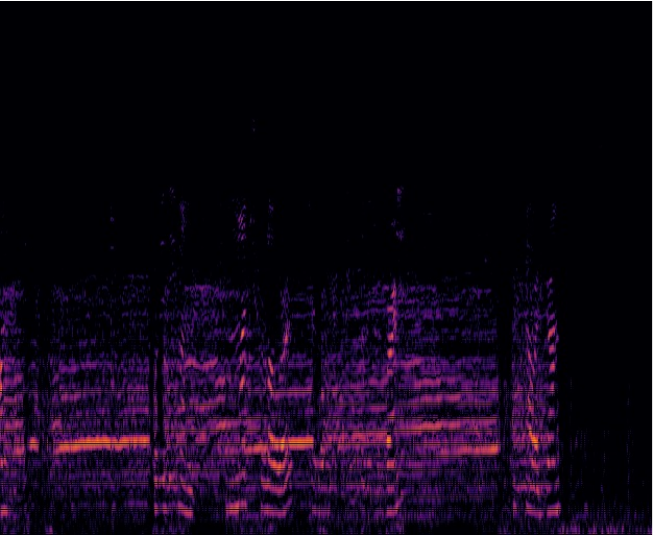}                      
&\includegraphics[width=0.18\textwidth]{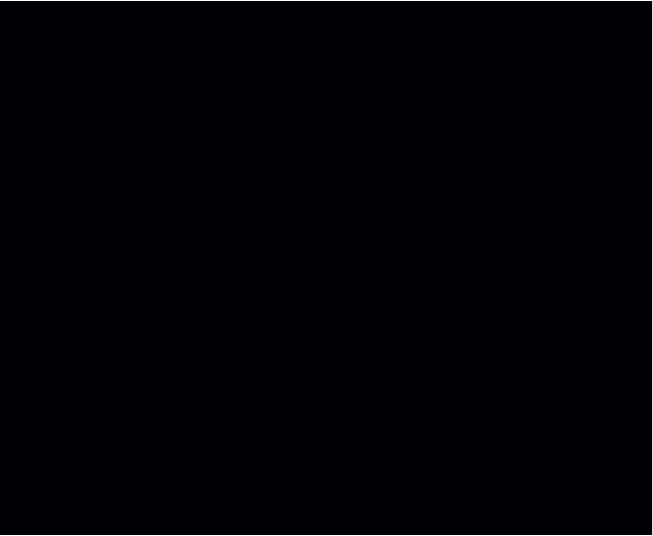}   \\ 
\bottomrule
\end{tabular}
\end{sc}
\end{small}
\end{center}
\vskip -0.2in
\end{table*}

\begin{table*}[]
\caption{Qualitative Results of Data Engine for Sound Separation. The two class labels serve as negative queries for each other during sound separation. Re-mixed audio refers to the audio obtained by mixing the two separated tracks back together.}
\label{tab:data_image}
\vskip 0.15in
\begin{center}
\begin{small}
\begin{sc}
\begin{tabular}{@{}ccccc@{}}
\toprule
\textbf{Query} & \textbf{Mixture} & \textbf{Track (Class1)} & \textbf{Track (Class2)} & \textbf{Re-mixed Audio}\\ \midrule
\includegraphics[width=0.18\textwidth]{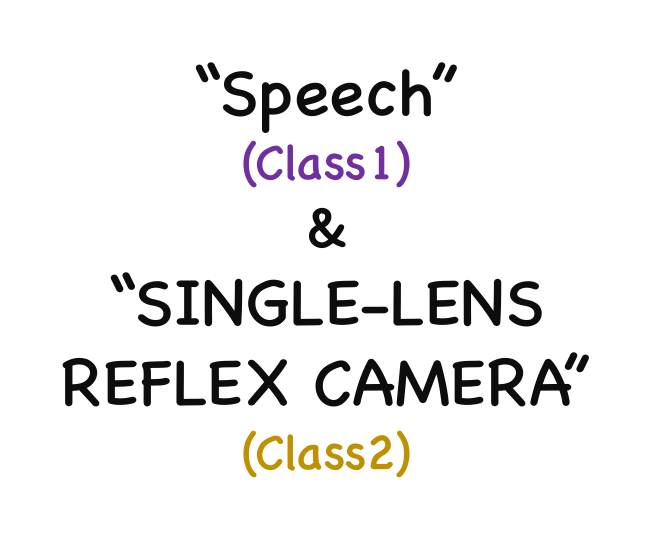}         
&\includegraphics[width=0.18\textwidth]{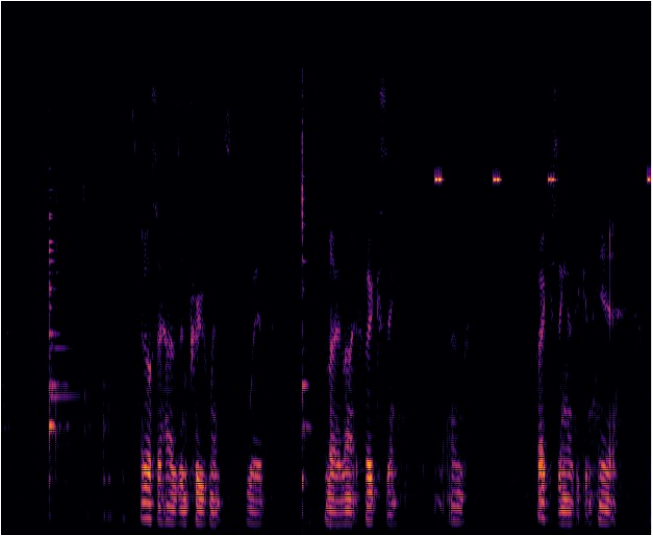}         
&\includegraphics[width=0.18\textwidth]{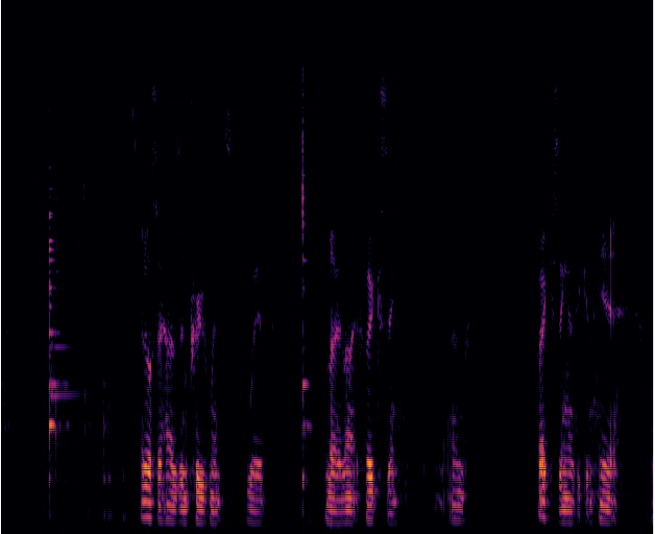}                   
&\includegraphics[width=0.18\textwidth]{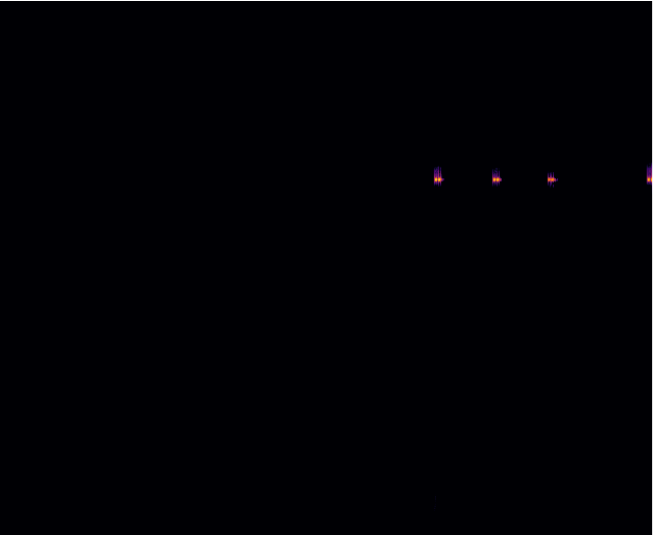}                      
&\includegraphics[width=0.18\textwidth]{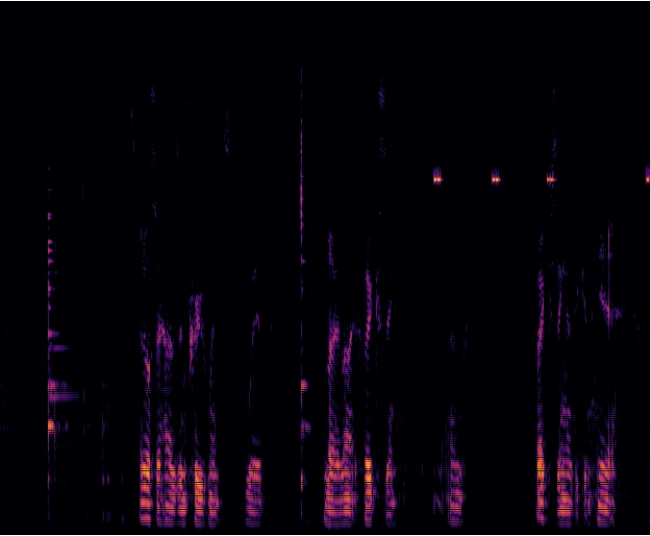}   \\ 
\bottomrule
\end{tabular}
\end{sc}
\end{small}
\end{center}
\vskip -0.1in
\end{table*}

\subsection{Qualitative Results}

Table~\ref{tab:silence_image} showcases the qualitative results of sound separation with non-existing queries. Both AudioSep and CLAPSep struggle to fully suppress interfering signals, leading to residual noise unrelated to the query. In contrast, ClearSep achieves significantly cleaner separation, producing near-silent outputs when the queried sound is missing. This demonstrates that our model ensures more precise separation, effectively preventing irrelevant noise from being retained. Moreover, this high-purity separation enables the extracted independent tracks to be suitable for further model training, underscoring the importance of silence augmentation in achieving data engine for sound separation.

Table~\ref{tab:data_image} further highlights the effectiveness of our data engine in refining sound separation quality. The example achieves an Re-SDR of 35.123, indicating that the separated tracks are well-isolated with minimal signal leakage. This result underscores the ability of our data engine to enhance separation purity and reduce interference between independent tracks.

Beyond these cases, we provide an extensive set of qualitative results on our demo page, including sound separation results on AudioCaps, additional silence separation examples, more data engine separation cases, and real-world sound separation results. We strongly encourage exploring the demo page to examine these cases in detail and listen to the separated audio samples.

\section{Conclusion}

Universal sound separation is crucial for audio understanding, yet existing methods heavily depend on artificially mixed data, limiting their effectiveness on naturally mixed audio. To address this, we propose ClearSep, a data engine framework that leverages naturally mixed audio for sound separation. We introduce Re-SDR and Re-SISDR, unsupervised metrics that evaluate separation reliability by comparing the remix of separated tracks with the original mixture. Using this approach, ClearSep decomposes mixed-source audio into high-quality single-source tracks, expanding the training dataset and enabling more advanced sound separation training. ClearSep operates through an iterative process, alternating between data augmentation via the data engine and training with the augmented data to progressively enhance sound separation performance. Additionally, we develop specialized training strategies to improve model robustness in real-world scenarios. 
Experimental results demonstrate that ClearSep achieves state-of-the-art performance, producing clean and reliable separations that are well-suited for further model fine-tuning. Benefiting from the purity of its separation results and our proposed unsupervised sound separation quality evaluation method, ClearSep ensures high-fidelity sound isolation while maintaining robustness across diverse real-world scenarios. Moving forward, ClearSep has the potential to become a powerful data processing tool, facilitating advancements in self-supervised sound separation, large-scale dataset curation.

\appendix


\section{More Implementation Details}
\label{app:implementation}

\subsection{More Training Details}
\label{app:training}
ClearSep adopts the CLAP model as its audio encoder, following the implementation in ~\citet{ma2024clapsep}, which achieves a zero-shot classification accuracy of 90.14\% on ESC-50. The model processes both linear spectrograms and Mel-spectrograms extracted from audio clips. The linear spectrogram is computed with a window length of 1024 points and a hop size of 320 points, resulting in 1001 frames with 513 frequency bins. The Mel-spectrogram follows the configuration of CLAP~\citep{laionclap2023}, where the audio is up-sampled to 48kHz, and a linear spectrogram is computed using a window length of 1024 points with a hop size of 480 points. This linear spectrogram is then transformed into 64 Mel bins, optimizing the representation for auditory perception analysis. The model architecture of ClearSep remains consistent with ~\citet{ma2024clapsep} in terms of layer configuration. Specifically, the encoder consists of 2, 2, 12, and 2 Transformer layers, the decoder includes 1, 1, 1, and 1 Transformer layers, and MaskNet comprises 3 Transformer layers. During training, only the separation decoder and LoRA modules applied to the CLAP audio encoder remain learnable, where LoRA is integrated into the query, key, value, and output projection layers of all multi-head attention modules with a rank of 16. The model is optimized using the AdamW optimizer with $\beta_1$ = 0.9, $\beta_2$ = 0.999, weight decay = $1e-2$, and a batch size of 256. The learning rate starts at $2e-4$ and decays exponentially to $5e-6$ with a factor of 0.3, triggered when the validation loss fails to improve for five consecutive epochs. We apply silence augmentation with a probability of $\alpha = 0.05$.

\vskip -0.1in
\begin{table}[h]
\caption{Data Statistics of the Data Engine. \textbf{\textsc{Thld.}} represents the threshold that both Re-SDR and Re-SISDR must simultaneously satisfy. \textbf{\textsc{Source}} refers to the mixed-source audio tracks being separated, while \textbf{\textsc{Separated}} denotes the separated independent tracks.}
\label{tab:engine_imple}
\begin{center}
\begin{small}
\begin{sc}
\begin{tabular}{@{}cc|rrrr@{}}
\toprule
\multirow{2}{*}{\textbf{Iter}} & \multirow{2}{*}{\textbf{Thld.}} & \multicolumn{2}{c}{\textbf{Source}}                      & \multicolumn{2}{c}{\textbf{Separated}}                     \\ \cmidrule(l){3-6} 
                      &                            & \multicolumn{1}{c}{\textbf{Num}} & \multicolumn{1}{c}{\textbf{Hour}} & \multicolumn{1}{c}{\textbf{Num}} & \multicolumn{1}{c}{\textbf{Hour}} \\ \midrule
Iter1                 & 10                         & 591K                    & 1 641h                       & 1 307K                  & 3 629h                       \\
Iter1                 & 15                         & 254K                    & 705h                         & 537K                    & 1 491h                       \\ \midrule
Iter2                 & 10                         & 730K                    & 2 028h                       & 1 616K                  & 4 488h                       \\
Iter2                 & 15                         & 559K                    & 1 551h                       & 1 187K                  & 3 297h                       \\ \bottomrule
\end{tabular}
\end{sc}
\end{small}
\end{center}
\vskip -0.2in
\end{table}

\subsection{More Details for Data Engine}
\label{app:data_engine}
During the data engine process, we use AudioSet labels~\citep{gemmeke2017audio} as queries for iterative refinement. To ensure mutual exclusivity among labels, we utilize the AudioSet Ontology to filter out parent labels that contain sub-labels within the same audio instance. This guarantees that each selected label represents a distinct sound category, reducing label overlap. In our framework, all labels serve as negative queries for other labels during sound separation, enforcing mutual exclusivity. This ensures that each separated track retains only the target label’s corresponding sound while filtering out all unrelated sources.

In Table~\ref{tab:engine_imple}, we provide a detailed statistical analysis of the separated audio tracks across different data engine iterations, illustrating the progressive dataset refinement and the expansion of high-quality training samples. For the sample to be included in SST, it must satisfy both Re-SDR$> 15\,$dB and Re-SISDR$> 15\,$dB, ensuring that only high-purity separations contribute to self-supervised learning. Meanwhile, samples used for ITT must meet a threshold of Re-SDR$> 10\,$dB and Re-SISDR$> 10\,$dB, allowing for a broader range of training data while maintaining sufficient separation quality. During the first iteration, we train the model for 80 epochs. In subsequent iterations, the model undergoes fine-tuning for 20 additional epochs, using the checkpoint from the previous iteration.

\section{More Evaluation Metrics Details.}

\subsection{General Metrics for Sound Separation}
Sound separation performance is typically evaluated using signal-to-distortion ratio (SDR) and scale-invariant signal-to-distortion ratio (SISDR), along with their respective improvements (SDRi and SISDRi). These metrics quantify the quality of separated audio by measuring its similarity to the ground truth source while penalizing distortions and interference.

\textbf{SDR (Signal-to-Distortion Ratio).} Signal-to-Distortion Ratio (SDR) evaluates the overall quality of the separated signal by measuring the ratio of the energy of the clean target signal to the energy of distortions, including noise and artifacts:
\begin{equation}
\text{SDR}(\hat{X}, X) = 10 \log_{10} \frac{|X|^2}{|X -\hat{X}|^2},
\end{equation}
where $X$ is the ground truth target signal, and $\hat{X}$ is the estimated separated signal.

\textbf{SDRi (SDR Improvement).}
SDRi quantifies the relative improvement achieved by a separation model compared to the original mixture signal. It is computed as the difference between the SDR of the separated signal and the SDR of the mixture signal:
\begin{equation}
\text{SDRi} = \text{SDR}(\hat{X}, X) - \text{SDR}(\tilde{X}, X),
\end{equation}
where  $\tilde{X}$ is the original mixed audio signal.

\textbf{SISDR (Scale-Invariant Signal-to-Distortion Ratio).} 
Scale-Invariant SDR (SISDR) addresses issues with SDR by removing the dependency on signal amplitude. It is computed by projecting the estimated signal onto the ground truth target and measuring the distortion in the residual:
\begin{equation}
\text{SISDR}(\hat{X}, X) = 10 \log_{10} \frac{| S_{\text{target}} |^2}{| \hat{X} - S_{\text{target}} |^2},
\end{equation}
where $S_{\text{target}} = \frac{\langle \hat{X}, X \rangle}{| X |^2} X$.
This ensures that the evaluation is independent of the absolute amplitude of the ground truth signal.

\textbf{SISDRi(SISDR Improvement).}
Similar to SDRi, SISDRi measures the improvement of SISDR after separation relative to the mixture:
\begin{equation}
\text{SISDRi} = \text{SISDR}(\hat{X}, X) - \text{SISDR}(\tilde{X}, X)
\end{equation}

\subsection{Sound Separation Purity Evaluation}
\label{app:silence}

Following prior work~\citep{wang2024consistent}, we adopt Silence-SDR and Silence-SISDR to evaluate whether a sound separation model effectively suppresses irrelevant audio when queried with a non-existing source. These metrics quantify the degree of silence preservation when the model is given a query label that does not correspond to any actual sound source in the input mixture. Given a mixed-source audio signal $X$, if the target query is absent from $X$, an ideal separation model should output complete silence, denoted as $X_{\text{silent}} = \mathcal{0}$. However, since $\mathcal{0}$ cannot be directly used as a target reference in standard SDR calculations, we evaluate separation purity by computing the Signal-to-Distortion Ratio (SDR) between the residual component $X-\hat{X}_{\text{silent}}$ and the original mixture $X$, where $\hat{X}_{\text{silent}}$ represents the model’s predicted output when queried with the absent source.

\textbf{Silence-SDR.} Silence-SDR assesses how much unintended residual noise remains in the predicted silent track $\hat{X}{\text{silent}}$, using the following formula:
\begin{equation}
\text{Silence-SDR}(X-\hat{X}_{\text{silent}}, X) = 10 \log_{10} \frac{| X |^2}{| \hat{X}_{\text{silent}} |^2}.
\end{equation}
A higher Silence-SDR indicates that the model effectively removes unwanted sounds, approaching an ideal silence.

\textbf{Silence-SISDR.} Similar to Silence-SDR, the Silence-Scale-Invariant SDR (Silence-SISDR) provides a scale-invariant evaluation by first computing the projection of the predicted silent output onto the original mixture:
\begin{small}
\begin{align}
\text{Silence-SISDR}(X\text{-}\hat{X}_{\text{silent}}, X) 
&\text{=}10 \log_{10} \frac{|S_{\text{silent}}|^2}
{| X\text{-}\hat{X}_{\text{silent}}\text{-}S_{\text{silent}}|^2},
\end{align}
\end{small}
where $S_{\text{silent}} = \frac{\langle X - \hat{X}_{\text{silent}}, X \rangle}{| X |^2} X$.

Both \textbf{Silence-SDR} and \textbf{Silence-SISDR} serve as critical indicators of sound separation quality. High values in these metrics indicate that the model successfully eliminates unintended audio when queried with a non-existing source, ensuring cleaner and more precise separation results.

\subsection{Remix-Based Unsupervised Evaluation}
\label{app:remix}
Traditional sound separation evaluation metrics such as SDR and SISDR require access to ground-truth isolated sources, which limits their applicability in real-world scenarios where such references are unavailable. To overcome this limitation, we propose \textbf{Re-SDR} and \textbf{Re-SISDR}, two remix-based unsupervised evaluation metrics that assess the quality of separated audio tracks without requiring ground-truth supervision.

Our approach is based on the principle that a well-separated set of independent tracks should be \textit{mutually exclusive} and \textit{collectively exhaustive}. This means that remixing the separated tracks should reconstruct the original mixed-source audio with minimal distortion. Given a mixed-source audio $\bar{X}$, our model separates it into multiple independent tracks $\{\hat{X}_1, \hat{X}_2, …, \hat{X}_n\}$. These separated tracks are then remixed to obtain the estimated mixture:
\begin{equation}
\bar{X} = \sum_{i=1}^{n} \hat{X}_i.
\end{equation}

To quantify the quality of sound separation, we compute the signal-to-distortion ratio (SDR) between the remixed audio $\bar{X}$ and the original mixed-source audio $X$:
\begin{equation}
\text{Re-SDR}(\bar{X}, X) = 10 \log_{10} \frac{\| X \|^2}{\| X - \bar{X} \|^2}.
\end{equation}

Furthermore, we define a scale-invariant version of remix-based SDR, termed \textbf{Re-SISDR}, which removes dependency on absolute signal amplitude by computing the projection of $\bar{X}$ onto $X$:
\begin{equation}
S_{\text{target}} = \frac{\langle \bar{X}, X \rangle}{\| X \|^2} X.
\end{equation}
Using this target estimate, the Re-SISDR is computed as:
\begin{equation}
\text{Re-SISDR}(\bar{X}, X) = 10 \log_{10} \frac{\| S_{\text{target}} \|^2}{\| \bar{X} - S_{\text{target}} \|^2}.
\end{equation}

These metrics provide a robust and unsupervised means of evaluating separation quality by leveraging remix consistency, making them suitable for real-world scenarios where ground-truth sources are unavailable.

\section{More Experiments}

\subsection{Single-Source Audio vs. Mixed-Source Audio}

As discussed earlier, many existing sound separation studies~\citep{liu2023separate,ma2024clapsep} incorporate substantial amounts of mixed-track data into their training pipelines to expand the scale of training data. However, this raises a critical question: Can training exclusively on clean single-track data lead to better performance? To investigate this, we conducted a controlled comparison, as summarized in Table~\ref{tab:tracks}. To ensure a fair evaluation, we maintained the same total number of tracks in each condition by creating mixed tracks through random pairing of clean tracks. This allowed us to directly compare models trained on mixed-track data versus those trained on clean single-track data.

\label{app:clean}
\begin{table}[t]
\caption{Comparison of models trained with different audio data: Single-Source Audios vs. Mixed-Source Audios.}
\label{tab:tracks}
\setlength\tabcolsep{4pt}
\vskip 0.15in
\begin{center}
\begin{small}
\begin{sc}
\begin{tabular}{@{}lcccc@{}}
\toprule
\textbf{Tracks} & \textbf{SDR}           & \textbf{SDRi}          & \textbf{SI-SDR}        & \textbf{SI-SDRi}       \\ \midrule
Mixed         & 8.73$\pm$7.18          & 8.72$\pm$7.19          & 7.82$\pm$7.67          & 7.82$\pm$7.67          \\
Single         & \textbf{9.18$\pm$7.41} & \textbf{9.16$\pm$7.41} & \textbf{8.29$\pm$7.91} & \textbf{8.29$\pm$7.91} \\ \bottomrule
\end{tabular}
\end{sc}
\end{small}
\end{center}
\vskip -0.2in
\end{table}

The results reveal that models trained on clean single-track data consistently outperform those trained on mixed-track data across various metrics, including SDR, SDRi, SI-SDR, and SI-SDRi. Notably, both the lower and upper performance bounds are higher when training on clean tracks. For instance, in terms of SDR, the separation upper bound for models trained with clean tracks reaches approximately 16.59 dB, which is 0.68 dB higher than the corresponding upper bound of 15.91 dB for models trained with mixed tracks. This discrepancy can be attributed to the interference between different sound events present in mixed-track data, which makes it more challenging for the model to distinguish and separate individual audio sources. For example, in the training data, dog barking is often accompanied by human voices, leading to scenarios where the model struggles to isolate the barking sound and inadvertently retains portions of human speech. In contrast, models trained on clean single-track data are better equipped to learn distinct representations of individual events, resulting in significantly clearer and more precise audio separation. These experimental findings underscore the substantial advantages of using high-quality, clean single-track data for training sound separation models. By decomposing mixed-source audio into single-source tracks for training, we effectively unlock the power within mixed audio data, enhancing the robustness and generalization of sound separation models.

\begin{table}[]
\caption{Comparison of sound separation models trained with different datasets on the AudioCaps. \textbf{P} and \textbf{N} denote positive and negative queries, respectively. Note that \textbf{AC} refers to training with AudioCaps, while \textbf{AS} refers to training with AudioSet.}
\label{tab:label}
\setlength\tabcolsep{2pt}
\begin{center}
\begin{small}
\begin{sc}
\begin{tabular}{@{}lllccc@{}}
\toprule
\multicolumn{1}{c}{}                         & \multicolumn{2}{l}{\textbf{Query Types}}                                & \multicolumn{1}{c}{}                             & \multicolumn{1}{c}{}                       & \multicolumn{1}{c}{}                          \\ \cmidrule(lr){2-3}
{\multirow{-2}{*}{\textbf{ID}}} & \multicolumn{1}{l}{\textbf{Train}} & \multicolumn{1}{l}{\textbf{Test}} & \multicolumn{1}{c}{\multirow{-2}{*}{\textbf{Scale}}} & \multicolumn{1}{c}{\multirow{-2}{*}{\textbf{SDRi}}} & \multicolumn{1}{c}{\multirow{-2}{*}{\textbf{SI-SDRi}}} \\ \midrule
\multicolumn{6}{l}{\cellcolor[HTML]{EFEFEF}\textit{Querying with Caption \textcolor{red}{(Requires extra effort to refine query.)}}}                                                                                                  \\
$\textit{E}_1$        & caption              & caption (P)                & AC                 & 9.27$\pm$5.19                     & 8.53$\pm$5.37                        \\ 
$\textit{E}_2$        & label              & caption (P)                & AS                 & \textbf{9.83$\pm$5.70}                     & \textbf{9.19$\pm$5.95}                        \\ 
\midrule
\multicolumn{6}{l}{\cellcolor[HTML]{EFEFEF}\textit{Querying with Class Labels \textcolor{customgreen}{(Requires no effort to refine query.)}}}                                                                                             \\
$\textit{E}_3$         & caption             & label (P)                & AC                 & 8.16$\pm$6.67                     & 7.26$\pm$7.06                        \\
$\textit{E}_4$      & label             & label (P)                & AC                 & 8.41$\pm$6.05                     & 7.63$\pm$6.32                        \\
$\textit{E}_5$     & label             & label (P+N)                & AC                 & 8.98$\pm$4.93                      & 8.24$\pm$5.01                        \\
$\textit{E}_6$      &label           & label (P)               & AS               & 9.18$\pm$6.77                      & 8.26$\pm$7.23                        \\
$\textit{E}_7$      & label           & label (P+N)               & AS               & 
\textbf{9.95$\pm$5.22}                      & \textbf{9.35$\pm$5.39}                                      \\ \bottomrule
\end{tabular}
\end{sc}
\end{small}
\end{center}
\vskip -0.1in
\end{table}

\subsection{Label-Queried vs. Caption-Queried}
\label{app:lab_cap}
Using fine-grained captions for sound separation allows the model to achieve precise audio track descriptions, while audio class labels are easier to obtain and can significantly expand the scale of training data. Each approach has its strengths and weaknesses. Therefore, we compared and revisited label-queried sound separation and caption-queried sound separation, as shown in Table \ref{tab:label}.

\textbf{Ease of Obtaining Queries.} An important aspect we examined is the ease of obtaining queries. As demonstrated by the comparison between $\textit{E}_1$ and $\textit{E}_3$, models trained with captions require high-quality queries that are sufficiently detailed and accurate. When captions lack precision, the performance can even fall behind that of models trained with labels ($\textit{E}_3$ vs. $\textit{E}_4$). This highlights the need for meticulous query refinement when using caption-queried sound separation models.

In contrast, label-queried sound separation models benefit from readily available labels generated by sound recognition systems. Since label categories are relatively fixed, they do not require additional optimization to function as queries. Moreover, different label categories can act as negative queries, further improving separation performance. While caption-queried models rely on extra query refinement to achieve optimal results, label-queried models leverage model-generated labels for automated sound separation, offering significantly greater convenience and ease of use.

\textbf{Scalability and Performance.} Although caption-queried models, with their fine-grained track descriptions, can achieve better sound separation performance on datasets of the same size (e.g., $\textit{E}_1$ vs. $\textit{E}_4$), audio-label datasets are far easier to obtain compared to audio-caption datasets. Leveraging larger training datasets, label-queried models (e.g., $\textit{E}_2$, $\textit{E}_7$) can easily surpass caption-queried models (e.g., $\textit{E}_1$). This demonstrates that label-queried sound separation methods are more scalable and better suited for utilizing large datasets to enhance separation performance.

\section*{Impact Statements}
Our ClearSep has the potential to significantly impact audio-centric applications, including speech enhancement, music separation, environmental sound analysis, and creation of multimedia content. By demonstrating that naturally mixed audio can serve as a rich and self-improving training resource, we pave the way for future advancements in self-supervised and weakly supervised sound separation. Furthermore, ClearSep achieves scaling without extensive human supervision, making it a promising tool for large-scale audio dataset curation.

However, our work also introduces potential risks related to the removal or alteration of certain sound information, which could be misused for unauthorized data manipulation or illicit synthetic audio generation. To mitigate such risks, we propose the integration of watermarking techniques within the separation pipeline, ensuring traceability, accountability, and regulatory oversight of generated audio content, thus safeguarding against unethical applications.

\bibliography{example_paper}

\begin{thebibliography}{32}
\providecommand{\natexlab}[1]{#1}
\providecommand{\url}[1]{\texttt{#1}}
\expandafter\ifx\csname urlstyle\endcsname\relax
  \providecommand{\doi}[1]{doi: #1}\else
  \providecommand{\doi}{doi: \begingroup \urlstyle{rm}\Url}\fi

\bibitem[Chen et~al.(2025)Chen, Ma, Li, Xu, Liang, Zheng, Yu, and Chen]{chen2025slam}
Chen, W., Ma, Z., Li, X., Xu, X., Liang, Y., Zheng, Z., Yu, K., and Chen, X.
\newblock Slam-aac: Enhancing audio captioning with paraphrasing augmentation and clap-refine through llms.
\newblock \emph{Proc. ICASSP}, 2025.

\bibitem[Cheng et~al.(2024)Cheng, Zheng, Wang, Fang, Zhang, Huang, Ma, Ji, Zuo, Jin, et~al.]{cheng2024omnisep}
Cheng, X., Zheng, S., Wang, Z., Fang, M., Zhang, Z., Huang, R., Ma, Z., Ji, S., Zuo, J., Jin, T., et~al.
\newblock Omnisep: Unified omni-modality sound separation with query-mixup.
\newblock \emph{arXiv preprint arXiv:2410.21269}, 2024.

\bibitem[D{\'e}fossez et~al.(2019)D{\'e}fossez, Usunier, Bottou, and Bach]{defossez2019demucs}
D{\'e}fossez, A., Usunier, N., Bottou, L., and Bach, F.
\newblock Demucs: Deep extractor for music sources with extra unlabeled data remixed.
\newblock \emph{arXiv preprint arXiv:1909.01174}, 2019.

\bibitem[Gemmeke et~al.(2017)Gemmeke, Ellis, Freedman, Jansen, Lawrence, Moore, Plakal, and Ritter]{gemmeke2017audio}
Gemmeke, J.~F., Ellis, D.~P., Freedman, D., Jansen, A., Lawrence, W., Moore, R.~C., Plakal, M., and Ritter, M.
\newblock Audio set: An ontology and human-labeled dataset for audio events.
\newblock In \emph{2017 IEEE international conference on acoustics, speech and signal processing (ICASSP)}, pp.\  776--780. IEEE, 2017.

\bibitem[Hershey et~al.(2016)Hershey, Chen, Le~Roux, and Watanabe]{hershey2016deep}
Hershey, J.~R., Chen, Z., Le~Roux, J., and Watanabe, S.
\newblock Deep clustering: Discriminative embeddings for segmentation and separation.
\newblock In \emph{2016 IEEE international conference on acoustics, speech and signal processing (ICASSP)}, pp.\  31--35. IEEE, 2016.

\bibitem[Hu et~al.(2021)Hu, Shen, Wallis, Allen-Zhu, Li, Wang, Wang, and Chen]{hu2021lora}
Hu, E.~J., Shen, Y., Wallis, P., Allen-Zhu, Z., Li, Y., Wang, S., Wang, L., and Chen, W.
\newblock Lora: Low-rank adaptation of large language models.
\newblock \emph{arXiv preprint arXiv:2106.09685}, 2021.

\bibitem[Kavalerov et~al.(2019)Kavalerov, Wisdom, Erdogan, Patton, Wilson, Le~Roux, and Hershey]{kavalerov2019universal}
Kavalerov, I., Wisdom, S., Erdogan, H., Patton, B., Wilson, K., Le~Roux, J., and Hershey, J.~R.
\newblock Universal sound separation.
\newblock In \emph{2019 IEEE Workshop on Applications of Signal Processing to Audio and Acoustics (WASPAA)}, pp.\  175--179. IEEE, 2019.

\bibitem[Kong et~al.(2020{\natexlab{a}})Kong, Cao, Iqbal, Wang, Wang, and Plumbley]{kong2020panns}
Kong, Q., Cao, Y., Iqbal, T., Wang, Y., Wang, W., and Plumbley, M.~D.
\newblock Panns: Large-scale pretrained audio neural networks for audio pattern recognition.
\newblock \emph{IEEE/ACM Transactions on Audio, Speech, and Language Processing}, 28:\penalty0 2880--2894, 2020{\natexlab{a}}.

\bibitem[Kong et~al.(2020{\natexlab{b}})Kong, Wang, Song, Cao, Wang, and Plumbley]{kong2020source}
Kong, Q., Wang, Y., Song, X., Cao, Y., Wang, W., and Plumbley, M.~D.
\newblock Source separation with weakly labelled data: An approach to computational auditory scene analysis.
\newblock In \emph{ICASSP 2020-2020 IEEE International Conference on Acoustics, Speech and Signal Processing (ICASSP)}, pp.\  101--105. IEEE, 2020{\natexlab{b}}.

\bibitem[Kristjansson et~al.(2006)Kristjansson, Hershey, Olsen, Rennie, and Gopinath]{kristjansson2006super}
Kristjansson, T.~T., Hershey, J.~R., Olsen, P.~A., Rennie, S.~J., and Gopinath, R.~A.
\newblock Super-human multi-talker speech recognition: the ibm 2006 speech separation challenge system.
\newblock In \emph{Interspeech}, volume~12, pp.\  155. Citeseer, 2006.

\bibitem[Le~Roux et~al.(2019)Le~Roux, Wisdom, Erdogan, and Hershey]{le2019sdr}
Le~Roux, J., Wisdom, S., Erdogan, H., and Hershey, J.~R.
\newblock Sdr--half-baked or well done?
\newblock In \emph{ICASSP 2019-2019 IEEE International Conference on Acoustics, Speech and Signal Processing (ICASSP)}, pp.\  626--630. IEEE, 2019.

\bibitem[Liu et~al.(2022)Liu, Liu, Kong, Mei, Zhao, Huang, Plumbley, and Wang]{liu2022separate}
Liu, X., Liu, H., Kong, Q., Mei, X., Zhao, J., Huang, Q., Plumbley, M.~D., and Wang, W.
\newblock Separate what you describe: Language-queried audio source separation.
\newblock \emph{arXiv preprint arXiv:2203.15147}, 2022.

\bibitem[Liu et~al.(2023)Liu, Kong, Zhao, Liu, Yuan, Liu, Xia, Wang, Plumbley, and Wang]{liu2023separate}
Liu, X., Kong, Q., Zhao, Y., Liu, H., Yuan, Y., Liu, Y., Xia, R., Wang, Y., Plumbley, M.~D., and Wang, W.
\newblock Separate anything you describe.
\newblock \emph{arXiv preprint arXiv:2308.05037}, 2023.

\bibitem[Luo \& Yu(2023)Luo and Yu]{luo2023music}
Luo, Y. and Yu, J.
\newblock Music source separation with band-split rnn.
\newblock \emph{IEEE/ACM Transactions on Audio, Speech, and Language Processing}, 2023.

\bibitem[Ma et~al.(2024)Ma, Peng, Shao, Liu, Li, and Wu]{ma2024clapsep}
Ma, H., Peng, Z., Shao, M., Liu, J., Li, X., and Wu, X.
\newblock Clapsep: Leveraging contrastive pre-trained models for multi-modal query-conditioned target sound extraction.
\newblock \emph{arXiv preprint arXiv:2402.17455}, 2024.

\bibitem[Maciejewski et~al.(2018)Maciejewski, Sell, Garcia-Perera, Watanabe, and Khudanpur]{maciejewski2018building}
Maciejewski, M., Sell, G., Garcia-Perera, L.~P., Watanabe, S., and Khudanpur, S.
\newblock Building corpora for single-channel speech separation across multiple domains.
\newblock \emph{arXiv preprint arXiv:1811.02641}, 2018.

\bibitem[Manilow et~al.(2019)Manilow, Wichern, Seetharaman, and Le~Roux]{manilow2019cutting}
Manilow, E., Wichern, G., Seetharaman, P., and Le~Roux, J.
\newblock Cutting music source separation some slakh: A dataset to study the impact of training data quality and quantity.
\newblock In \emph{2019 IEEE Workshop on Applications of Signal Processing to Audio and Acoustics (WASPAA)}, pp.\  45--49. IEEE, 2019.

\bibitem[Manilow et~al.(2022)Manilow, O’Reilly, Seetharaman, and Pardo]{manilow2022source}
Manilow, E., O’Reilly, P., Seetharaman, P., and Pardo, B.
\newblock Source separation by steering pretrained music models.
\newblock In \emph{ICASSP 2022-2022 IEEE International Conference on Acoustics, Speech and Signal Processing (ICASSP)}, pp.\  126--130. IEEE, 2022.

\bibitem[Ochiai et~al.(2020)Ochiai, Delcroix, Koizumi, Ito, Kinoshita, and Araki]{ochiai2020listen}
Ochiai, T., Delcroix, M., Koizumi, Y., Ito, H., Kinoshita, K., and Araki, S.
\newblock Listen to what you want: Neural network-based universal sound selector.
\newblock \emph{arXiv preprint arXiv:2006.05712}, 2020.

\bibitem[Pegg et~al.(2023)Pegg, Li, and Hu]{pegg2023rtfs}
Pegg, S., Li, K., and Hu, X.
\newblock Rtfs-net: Recurrent time-frequency modelling for efficient audio-visual speech separation.
\newblock \emph{arXiv preprint arXiv:2309.17189}, 2023.

\bibitem[Piczak(2015)]{piczak2015esc}
Piczak, K.~J.
\newblock Esc: Dataset for environmental sound classification.
\newblock In \emph{Proceedings of the 23rd ACM international conference on Multimedia}, pp.\  1015--1018, 2015.

\bibitem[Pons et~al.(2024)Pons, Liu, Pascual, and Serr{\`a}]{pons2024gass}
Pons, J., Liu, X., Pascual, S., and Serr{\`a}, J.
\newblock Gass: Generalizing audio source separation with large-scale data.
\newblock In \emph{ICASSP 2024-2024 IEEE International Conference on Acoustics, Speech and Signal Processing (ICASSP)}, pp.\  546--550. IEEE, 2024.

\bibitem[Roweis(2000)]{roweis2000one}
Roweis, S.
\newblock One microphone source separation.
\newblock \emph{Advances in neural information processing systems}, 13, 2000.

\bibitem[Smaragdis(2004)]{smaragdis2004non}
Smaragdis, P.
\newblock Non-negative matrix factor deconvolution; extraction of multiple sound sources from monophonic inputs.
\newblock In \emph{Independent Component Analysis and Blind Signal Separation: Fifth International Conference, ICA 2004, Granada, Spain, September 22-24, 2004. Proceedings 5}, pp.\  494--499. Springer, 2004.

\bibitem[Vincent et~al.(2006)Vincent, Gribonval, and F{\'e}votte]{vincent2006performance}
Vincent, E., Gribonval, R., and F{\'e}votte, C.
\newblock Performance measurement in blind audio source separation.
\newblock \emph{IEEE transactions on audio, speech, and language processing}, 14\penalty0 (4):\penalty0 1462--1469, 2006.

\bibitem[Wang \& Chen(2018{\natexlab{a}})Wang and Chen]{Wang_Chen_2018}
Wang, D. and Chen, J.
\newblock Supervised speech separation based on deep learning: An overview.
\newblock \emph{IEEE/ACM Transactions on Audio, Speech, and Language Processing}, pp.\  1702–1726, Oct 2018{\natexlab{a}}.
\newblock \doi{10.1109/taslp.2018.2842159}.
\newblock URL \url{http://dx.doi.org/10.1109/taslp.2018.2842159}.

\bibitem[Wang \& Chen(2018{\natexlab{b}})Wang and Chen]{wang2018supervised}
Wang, D. and Chen, J.
\newblock Supervised speech separation based on deep learning: An overview.
\newblock \emph{IEEE/ACM transactions on audio, speech, and language processing}, 26\penalty0 (10):\penalty0 1702--1726, 2018{\natexlab{b}}.

\bibitem[Wang et~al.(2024)Wang, Chen, Yang, Yu, Weng, Wu, and Meng]{wang2024consistent}
Wang, Y., Chen, H., Yang, D., Yu, J., Weng, C., Wu, Z., and Meng, H.
\newblock Consistent and relevant: Rethink the query embedding in general sound separation.
\newblock In \emph{ICASSP 2024-2024 IEEE International Conference on Acoustics, Speech and Signal Processing (ICASSP)}, pp.\  961--965. IEEE, 2024.

\bibitem[Wang et~al.(2023)Wang, Cornell, Choi, Lee, Kim, and Watanabe]{wang2023tf}
Wang, Z.-Q., Cornell, S., Choi, S., Lee, Y., Kim, B.-Y., and Watanabe, S.
\newblock Tf-gridnet: Making time-frequency domain models great again for monaural speaker separation.
\newblock In \emph{ICASSP 2023-2023 IEEE International Conference on Acoustics, Speech and Signal Processing (ICASSP)}, pp.\  1--5. IEEE, 2023.

\bibitem[Wisdom et~al.(2020)Wisdom, Tzinis, Erdogan, Weiss, Wilson, and Hershey]{wisdom2020unsupervised}
Wisdom, S., Tzinis, E., Erdogan, H., Weiss, R., Wilson, K., and Hershey, J.
\newblock Unsupervised sound separation using mixture invariant training.
\newblock \emph{Advances in Neural Information Processing Systems}, 33:\penalty0 3846--3857, 2020.

\bibitem[Wu et~al.(2023)Wu, Chen, Zhang, Hui, Berg-Kirkpatrick, and Dubnov]{laionclap2023}
Wu, Y., Chen, K., Zhang, T., Hui, Y., Berg-Kirkpatrick, T., and Dubnov, S.
\newblock Large-scale contrastive language-audio pretraining with feature fusion and keyword-to-caption augmentation.
\newblock In \emph{IEEE International Conference on Acoustics, Speech and Signal Processing, ICASSP}, 2023.

\bibitem[Yu et~al.(2017)Yu, Kolb{\ae}k, Tan, and Jensen]{yu2017permutation}
Yu, D., Kolb{\ae}k, M., Tan, Z.-H., and Jensen, J.
\newblock Permutation invariant training of deep models for speaker-independent multi-talker speech separation.
\newblock In \emph{2017 IEEE International Conference on Acoustics, Speech and Signal Processing (ICASSP)}, pp.\  241--245. IEEE, 2017.

\end{thebibliography}
\bibliographystyle{icml2025}

\end{document}